\documentclass[12pt]{iopart}

\usepackage{amstext}
\usepackage{amssymb,amsthm} 
\usepackage{fullpage}
\usepackage{rotating}
\usepackage{longtable}
\usepackage{lscape}
\usepackage{url}

\newcommand{\be}{\begin{equation}}
\newcommand{\ee}{\end{equation}}
\newcommand{\ba}{\begin{eqnarray}}
\newcommand{\ea}{\end{eqnarray}}
\newcommand{\ket}[1]{\mbox{$ | #1 \rangle $}}
\newcommand{\moy}[1]{\langle #1 \rangle}

\begin{document}

\title{Extremal correlations of the\\ tripartite no-signaling polytope}

\author{Stefano Pironio$^1$, Jean-Daniel Bancal$^2$, 
Valerio Scarani$^{3}$}

\address{$^1$ Laboratoire d'Information Quantique, Universit\'e Libre de Bruxelles, 1050 Brussels, Belgium}
\address{$^2$ Group of Applied Physics, University of Geneva, Switzerland}
\address{$^3$ Centre for Quantum Technologies and Department of Physics, National University of Singapore, Singapore 117543}

\begin{abstract}
The no-signaling polytope associated to a Bell scenario with three parties, two inputs, and two outputs is found to have 53856 extremal points, belonging to 46 inequivalent classes. We provide a classification of these points according to various definitions of multipartite non-locality and briefly discuss other issues like the interconversion between extremal points seen as a resource and the relation of the extremal points to Bell-type inequalities. 
\end{abstract}
\maketitle

\normalsize

\section{Introduction}
Quantum correlations, i.e., probability distributions characterizing the outcomes of measurements performed on entangled quantum states, belong to the set of \textit{no-signaling probability distributions}. No-signaling captures one of the essential properties of quantum correlations: the impossibility of using them to send a message. Popescu and Rohrlich surmised that this property may define quantum correlations exactly; upon studying the question, however, they realized that it is not so and showed a probability distribution that satisfies no-signaling but cannot be obtained from quantum physics. This mathematical object is known as a  \textit{PR-box} \cite{pr}, although other authors had discussed it earlier \cite{rastall,tsi}.

In recent years, various authors have studied no-signaling distributions, their motivations ranging from sheer mathematical interest to the hope of describing something that may be discovered in nature. At any rate, the set of no-signaling probability distributions provides a thinking space \cite{blmppr}, in which one can meaningfully ask why nature is not more non-local \cite{popnatphys}, what are the physical principles that underlie quantum physics \cite{cc,ic,ml,ow}, or how to exploit nonlocality for information processing \cite{BHK,AGM}.

By treating each probability distribution as a point in a high-dimensional space, one obtains a geometric characterization of the set of no-signaling probability distributions: this set is a \textit{polytope}, i.e., a convex set with finitely many extremal points. The first no-signaling polytope to be characterized is associated to the simplest meaningful scenario: two parties, each with two inputs (the measurement settings) and two outputs (the measurement outcomes) \cite{tsi,blmppr}. This polytope lives in an 8-dimensional space. It has 24 vertices, 16 of which describe local deterministic distributions, while the 8 non-local points are all equivalent to the PR-box under suitable relabeling of the inputs and outputs. The facets of the local polytope also belong to two classes upon relabeling: 16 of them are positivity inequalities enforcing the constraint that probabilities must lie between 0 and 1; and there are 8 non-trivial facets, all equivalent to the Clauser-Horne-Shimony-Holt (CHSH) Bell-type inequality \cite{chsh}. The elegance of the construction is completed when one notices that each of the 8 non-local points lies above one of the 8 non-trivial facets of the local polytope.

In the past few years, other no-signaling polytopes for bipartite scenarios have been studied, namely: two inputs and $d$ outputs \cite{blmppr}, and $m$ inputs and two outputs \cite{bp05,jones}. In these examples, it was possible to give a compact description of the geometry of the polytopes, notwithstanding their growing complexity. The structure of no-signaling theories in \textit{multipartite scenarios}, on the contrary, has been only partially addressed in some of the initial studies \cite{blmppr,bp05}. Recent results motivate the need for a better understanding \cite{gyni}.

In this paper, we present the no-signaling polytope for \textit{three parties, two inputs, and two outputs} and derive its extremal points. This is the simplest multipartite scenario: nevertheless, the complexity of the geometry of the polytope is far greater than in the bipartite case. We discuss some possibilities for classifying the extremal boxes, being aware though that many questions remain open. In Section \ref{sec:boxes}, we define the mathematical objects, provide the list of extremal tripartite boxes, and discuss a few simple examples. In Section \ref{sec:nonloc}, we sketch several criteria for multipartite nonlocality and classify the extremal boxes according to these. In Section~\ref{sec:inter}, we briefly mention some known results on simulating some boxes using other ones. In Section~\ref{sec:bell}, we study the violation of Bell's inequalities by the extremal boxes.

\section{Tripartite no-signaling boxes}
\label{sec:boxes}

\subsection{Notation and definitions}
We are interested in the set of tripartite no-signaling boxes, where each party has  two inputs and two outputs. Let $x,y,z\in\{0,1\}$ denote the inputs of each party and $a,b,c\in\{-1,1\}$, the outputs. The boxes are characterized by the joint probabilities $P(abc|xyz)$ of obtaining the triple of outputs $(a,b,c)$ given the triple of inputs $(x,y,z)$.	These probabilities satisfy positivity
\be \label{pos}
P(abc|xyz)\geq 0,\,\quad \text{for all $a,b,c,x,y,z$,}
\ee
normalization
\be \label{norma}
\sum_{a,b,c} P(abc|xyz)=1,\,\quad \text{for all $x,y,z$,}
\ee
and no-signaling
\be \label{nosig}
\sum_{c} P(abc|xyz)=\sum_{c'} P(abc'|xyz'),\,\quad \text{for all $a,b,x,y,z,z'$,}
\ee
where the last condition also holds for cyclic permutations of the parties. These no-signaling conditions guarantee that signaling among any partitions of the parties is impossible, e.g., that box C cannot signal to boxes A or B, or to the combined system AB, or conversely that the combined system AB cannot signal to C.

Due to the equality constraints (\ref{norma}) and (\ref{nosig}), only 26 out of the 64 probabilities $P(abc|xyz)$ are independent, i.e., the set of no-signaling boxes is contained in an affine space of dimension 26. It is convenient to write the no-signaling boxes in this 26-dimensional space using the following parametrisation 
\ba 
P(abc|xyz)&=&\frac{1}{8}\Big[1+ a \langle A_x\rangle +  b \langle B_y\rangle + c \langle C_z\rangle \nonumber\\ &&+  a b \langle A_x B_y\rangle + a c \langle A_xC_z\rangle +  b c \langle B_yC_z\rangle +  a b c \langle A_xB_yC_z\rangle\Big]\,,\ea
where $\langle A_x\rangle=P(a=1|x)-P(a=-1|x)$ is the expectation value of the outcome $a$ for the input $x$,  $\langle A_xB_y\rangle= P(ab=1|xy)-P(ab=-1|xy)$ is the expectation value of the product $ab$ for the inputs $x$ and $y$, and so on. Note that the single-party and two-party expectations are well-defined and do not depend on the other parties inputs (e.g., $\langle A_x\rangle = \langle A_{xyz}\rangle$) thanks to the no-signaling conditions. In total, there are 6 single-party expectations, 12 two-party expectations, and 8 three-party expectations, adding up to a total of 26 numbers that fully specify a probability point in the no-signaling set. 

It is sometimes useful to consider the ``computer scientist'' notation where outputs take values in $\{0,1\}$, instead of the ``physicist'' notation where they take value in  $\{-1,1\}$. We will therefore also consider the alternative labeling $\hat a,\hat b,\hat c\in \{0,1\}$ for the outputs, with $a=(-1)^{\hat a}$, $b=(-1)^{\hat b}$, and $c=(-1)^{\hat c}$.  With this labeling, it is convenient to parametrize a no-signaling point through the ``sum modulo 2" expectations $\langle \hat A_x\rangle = \sum_{\hat{a}}P(\hat{a}|x)\,\hat{a} $, $\langle \hat A_x+\hat B_y\rangle =\sum_{\hat{a},\hat{b}} P(\hat{a}\hat{b}|xy)\,(\hat{a}+\hat{b})$, and $\langle \hat A_x +\hat B_y +\hat C_z\rangle = \sum_{\hat{a},\hat{b},\hat{c}} P(\hat{a}\hat{b}\hat{c}|xyz)\,(\hat{a}+\hat{b}+\hat{c})$. These expectations are in one-to-one correspondence with the ``product'' expectations defined above through $\langle A_x\rangle=1-2\langle \hat A_x\rangle$, $\langle A_xB_y\rangle=1-2\langle \hat A_x+\hat B_y\rangle$, $\langle A_xB_yC_z\rangle=1-2\langle \hat A_x+\hat B_y+\hat C_z\rangle$. As an illustration, the PR box is defined in the physicist notation by $\langle A_x\rangle=0$, $\langle B_y\rangle=0$, $\langle A_xB_y\rangle=(-1)^{xy}$, and in the computer scientist notation by $\langle \hat A_x\rangle=1/2$, $\langle \hat B_y\rangle=1/2$, $\langle \hat{A}_x+\hat{B}_y\rangle=xy$.

\subsection{Extremal boxes}

Since the constraints (\ref{pos}), (\ref{norma}), and (\ref{nosig}) are linear, the set of no-signaling boxes is a polytope. Boxes of particular interest are the extremal ones, which correspond to the vertices of this polytope. They fully characterize the no-signaling polytope since any box can be decomposed as a convex combination of the extremal ones. Given a polytope described in term of linear constraints, there exist algorithms that can enumerate all its vertices, although they are efficient only for low dimensional problems.

We determined the extreme boxes of the the tripartite no-signaling polytope using both the algorithms \emph{PORTA} \cite{porta} and \emph{cdd} \cite{cdd}. It turns out that there are 53856 extremal points. These points can be classified by equivalence classes under relabeling of the parties, inputs, and outputs\footnote[7]{The relabeling of the inputs and outputs must be defined by a \textit{local} processing. For instance, $\hat{a}\rightarrow \hat{a}+x$ (sum modulo 2) is allowed; $\hat{a}\rightarrow \hat{a}+y$ is not.}. Once the extremal points are sorted according to these symmetries, they define 46 equivalence classes. We provide a representative for each class both in the physicist and computer scientist notation (Tables~\ref{tvertphys} and \ref{tvertcomp}). These lists are also available in electronic format \cite{website}.

\subsection{More detailed presentation of some boxes}
Let us start by featuring some extremal boxes of particular interest or which have already appeared in the literature.

\begin{itemize}
\item \emph{Deterministic boxes (boxes of class 1)}. Boxes in this category have deterministic outputs, for instance the representative provided in Tables 1 and 2 satisfies
\be 
\hat a_x =0,\quad \hat b_y =0,\quad  \hat c_z=0\,.
\ee
The 64 possible deterministic boxes define the extremal points of the polytope of local correlations. All the other extremal boxes are nonlocal.

\item \emph{PR boxes (boxes 2)}.  This class comprises boxes corresponding to a PR-box shared between two parties, with the third party deterministic. For instance, the representative 2 in Tables 1 and 2 satisfies the relations
\be 
\hat a_x=0,\quad \hat b_y+\hat c_z=yz\,.
\ee
These are in essence bipartite boxes. 

\item \emph{GYNI boxes (boxes 25, 29)}. These are the two no-signaling boxes associated to the tripartite ``guess your neighbor inputs'' (GYNI) non-local game \cite{gyni}. GYNI is a non-local game whose winning probability corresponds to the Bell expression
\be \label{gyniineq}
w=\frac{1}{4}\left[P(000|000)+P(110|011)+P(011|101)+P(101|110)\right].
\ee
Quantum correlations achieve at most $w=1/4$, which is not better than classical strategies. No-signaling correlations, however, can outperform classical and quantum strategies and achieve $w=1/3$. Boxes in the classes 25 and 29 are the two boxes achieving the maximum no-signaling winning probability $w=1/3$. 

\item \emph{Full-correlation boxes (boxes 44, 45, 46)}. These boxes are the only full-correlation boxes, for which all one-party and two-party correlation terms vanish. They can thus be written as 
\be \label{fullcorr}
P(abc|xyz)=\frac{1}{8}\left[1+abc\langle A_xB_yC_z\rangle\right]\,,
\ee
with
\be 
\langle A_xB_yC_z\rangle = (-1)^{xyz}\quad\text{for box 44},
\ee
\be 
\langle A_xB_yC_z\rangle = (-1)^{x(y+z)}\quad\text{for box 45},
\ee
\be \label{46}
\langle A_xB_yC_z\rangle = (-1)^{xy+xz+yz}\quad\text{for box 46}.
\ee
These boxes correspond to a situation with perfect correlations: for instance in the case of box 44, the outcomes satisfy $abc=-1$ if all parties use measurement ``1'', and they satisfy $abc=+1$ in all other cases. These are the only genuine tripartite boxes with this property.

These three boxes were already introduced in \cite{blmppr}. Boxes 46 were called ``Svetlichny'' boxes because they violate Svetlichny's original inequality \cite{svetlichny}
\be 
S=\sum_{xyz} (-1)^{xy+xz+yz}\langle A_xB_yC_Z\rangle\leq 4
\ee
up to the algebraic maximum $S=8$. Boxes 44 and 46 also violate the Mermin inequality~\cite{mermin}
\ba
M_3&=&\moy{A_1B_0C_0}+\moy{A_0B_1C_0}+\moy{A_0B_0C_1}-\moy{A_1B_1C_1}\,\leq\, 2\label{merminref}
\ea
up to its algebraic maximum $M_3=4$.
\end{itemize}

Note that it is also possible to violate maximally the Mermin inequality using quantum systems. By measuring the Greenberger-Horne-Zeilinger (GHZ) state $\frac{1}{\sqrt{2}}(\ket{000}+\ket{111})$ in suitable local bases  \cite{ghz}, one obtains correlations of the form $P_{GHZ}(abc|xyz)=\frac{1}{8}\left[1+abc\langle A_xB_yC_z\rangle\right]$ with
\ba
\moy{A_1B_0C_0}\,=\,\moy{A_0B_1C_0}\,=\,\moy{A_0B_0C_1}\,=\,-\moy{A_1B_1C_1}&=&1\,,\label{ghz1}\\
\moy{A_0B_0C_0}\,=\,\moy{A_0B_1C_1}\,=\,\moy{A_1B_0C_1}\,=\,\moy{A_1B_1C_0}&=&0\,.\label{ghz2}
\ea 
These correlations return $M_3=4$ for the the Mermin inequality. The quantum point $P_{GHZ}$ is not extremal, however; but it can be decomposed using extremal points in a very simple way:
\ba
P_{GHZ}&=&\frac{1}{2}\,P_{46}\,+\,\frac{1}{2}\,P'_{46}\,.
\ea
Here, $P_{46}$ is the extremal point given by (\ref{46}), while $P'_{46}$ is another point in the same class. $P_{46}'$ is defined by $\langle A_xB_yC_z\rangle=(-1)^{1+x+y+z+xy+xz+yz}$ and can be obtained from $P_{46}$ by the following local relabelling of the outputs: $a\rightarrow (-1)^{1-x}a,\,b\rightarrow (-1)^{1-y}b,\,c\rightarrow (-1)^{1-z}c$, i.e., in words, the parties flip their output for the input ``0" and leave it unchanged for the input ``1".
It can be checked that $P_{46}$ and $P'_{46}$ are the only two extremal points of the 46th class that reach $M_3=4$ for the representative Mermin inequality (\ref{merminref}). More generally, for each version of the Mermin inequality, there are 118 extremal points that reach $M_3=4$: 2 in the class 46 (as we have seen), 8 in the class 44, 12 in the class 2, and 32 in each of the classes 21, 22, and 34. The GHZ correlations defined by (\ref{ghz1}), (\ref{ghz2}) can be reproduced by mixing these strategies in an uniform way within each of these classes.

Before embarking in further characterizations of the extremal boxes, let us make some additional observations. Out of  the 46 types of extremal boxes, only 13 of them have their correlation terms that are either perfect or uniformly random, i.e., only 13 boxes have correlation terms that only take as possible values 0, 1, or -1. These are the boxes $1-8$, 41, 42, and $44-46$. Finally, the boxes which are the most symmetric under relabelling of parties, inputs, and outputs are boxes 46, with only 16 representatives in their equivalence class, while the least symmetric are boxes 14, 17, 32, 35, 37, 38, with 3072 different representatives.

\section{Classification through nonlocality}
\label{sec:nonloc}

As a first attempt at classifying these 46 different boxes, let us consider their basic non-local properties. Multipartite nonlocality is more intricate than the bipartite case, so we start by defining several possible criteria.

\subsection{Notions of multipartite nonlocality}

It is helpful to think about nonlocality in operational terms and ask what type of classical resources (shared randomness, communication) are needed by classical observers to simulate a particular kind of nonlocal box.

A box is said to be \emph{local} if it can be simulated by non-communicating classical observers using shared randomness only. A local box thus admits a decomposition of the form
\be \label{local}
P(abc|xyz)=\sum_\lambda q_{\lambda} P_\lambda(a|x)P_\lambda(b|y)P_\lambda(c|z)\,,
\ee
where the variable $\lambda$ has probability distribution $q_\lambda$ and can be thought of as the shared randomness determining the local response of each party. We denote $L$ the set of local boxes.
A box is said to be \emph{nonlocal} if it cannot be written in the above way, which implies that some communication between the parties is required to simulate it.

Among nonlocal boxes, one can distinguish further between those that require for their simulation only communication between two of the parties and those that require all three parties to communicate. Following Svetlichny's original definition \cite{svetlichny}, we thus say that a box is \emph{2-way Svetlichny nonlocal} if it admits a decomposition of the form
\be \label{svet}
P(abc|xyz)=q_1P^{AB/C}(abc|xyz)+q_2P^{AC/B}(abc|xyz)+q_3P^{BC/A}(abc|xyz)
\ee
where
\be \label{AB/C}
P^{AB/C}(abc|xyz)=\sum_\lambda q_{\lambda} P_\lambda(ab|xy)P_\lambda(c|z)
\ee
corresponds to a nonlocal term involving communication only between parties $A$ and $B$, and where $P^{AC/B}(abc|xyz)$ and $P^{BC/A}(abc|xyz)$ are similarly defined. We denote $S_2$ the set of 2-way Svetlichny  nonlocal boxes. A box is said to be 3-way Svetlichny nonlocal if it cannot be written in the above form.

Classical communication models \`a la Svetlichny presuppose that all parties receive their inputs at the same time\footnote[7]{The following discussion is a concise summary of a forthcoming paper on the definition of genuine multipartite non-locality \cite{inprep}. In particular, it will be argued in \cite{inprep} that the proper notion of genuine tripartite nonlocality should be based on \textit{US} nonlocality, see definition below, rather than Svetlichny's one.}. This is followed by one or several rounds of communication after which all parties produce an output. Inputs into no-signaling boxes need not, however, be given simultaneously to all parties. For instance, in quantum theory measurements on an entangled state can be performed in a sequence or on a subset of systems only. When one subsystem is measured, the outcome is obtained immediately, and one does not have to wait until all the other subsystems have also been measured. In analogy with the quantum case, the same feature can be thought of no-signaling boxes: once a party introduces an input into a box, an output is obtained immediately, irrespective of whether inputs have been introduced by other parties. This is possible thanks to the no-signaling condition which ensures that the marginal output probability distribution for any subset of the parties is well-defined and is independent of the inputs for the other parties.

When inputs are given in a sequence rather than simultaneously, it is necessary to consider communication models more restricted than Svetlichny's ones, since a party's output can depend on communications already received, but cannot depend on communications from parties later in the sequence. For instance, the nonlocal term $P_\lambda(ab|xy)$ in Eq.~(\ref{AB/C}) should be replaced by $P_\lambda(a|x)P_\lambda(b|xy)$ if party $A$ receives his inputs before party $B$, and by $P_\lambda(a|xy)P_\lambda(b|y)$ if it is party $B$ who receives it first.

We consider here two alternative notions of 2-way nonlocality based on such communication models with inputs given in a sequence. In the first model, inputs are given according to an arbitrary sequence which is known beforehand by the parties. We say that a box is \emph{2-way KS nonlocal} (where \textit{KS} stands for ``known sequence") if it can be reproduced using communication between at most two of the parties, irrespective of the predetermined input sequence. It is easy to show that a box is 2-way \textit{KS} nonlocal if it admits a decomposition of the form (\ref{svet}) where
\begin{eqnarray} 
P^{AB/C}(abc|xyz)&=&\sum_\mu q_{\mu} P_\mu(a|x)P_\mu(b|xy)P_\mu(c|z)\label{KSa}\\
&=&\sum_\nu q_{\nu} P_\nu(a|xy)P_\nu(b|y)P_\nu(c|z)\label{KSb}
\end{eqnarray}
and similarly for the terms $P^{AC/B}(abc|xyz)$ and $P^{BC/A}(abc|xyz)$. Eq. (\ref{KSa}) specifies the response of the parties when $A$ precedes $B$ in the sequence and Eq. (\ref{KSb}) when it is $B$ who precedes $A$. The response of the parties in each case is determined, respectively, by different set of random variables $\{\mu\}$ and $\{\nu\}$. The set of 2-way \textit{KS} nonlocal boxes is denoted $\mathit{KS}_2$.

In the second model, inputs are given according to an arbitrary sequence which is not known in advance by the parties and we say that a box is \emph{2-way US nonlocal} (where \textit{US} stands for ``unkown sequence") if it can be reproduced using communication between at most two of the parties, irrespective of the unknown input sequence. A box is 2-way \textit{US} nonlocal if it admits a decomposition of the form (\ref{svet}) with
\begin{eqnarray} 
P^{AB/C}(abc|xyz)&=&\sum_{\lambda\mu} q_{\lambda}q_{\mu|\lambda} P_\mu(a|x)P_\mu(b|xy)P_\lambda(c|z)\label{USa}\\
&=&\sum_{\lambda\nu} q_{\lambda}q_{\lambda|\nu} P_\nu(a|xy)P_\nu(b|y)P_\lambda(c|z)\label{USb}
\end{eqnarray}
and similarly for the terms $P^{AC/B}(abc|xyz)$ and $P^{BC/A}(abc|xyz)$. As before, Eq. (\ref{USa}) specifies the response of the parties when $A$ precedes $B$ in the sequence and Eq. (\ref{USb}) when it is $B$ who precedes $A$. In each case, the behaviour of the parties is specified by two sets of shared variables $\{\lambda,\mu\}$ and $\{\lambda,\nu\}$. The difference with the previous definition is that party $C$ has no way to know the relative order between $A$ and $B$ as the input sequence is not specified in advance and as it is not communicating with the other parties.  The response of party $C$ is thus identical in each case and depends only on a variable $\lambda$ common to both sets $\{\lambda,\mu\}$ and $\{\lambda,\nu\}$. We denote $\mathit{US}_2$ the set of 2-way \textit{US} nonlocal boxes.

Finally, we also consider models where parties are allowed to use other no-signaling boxes as a resource rather than communication. We define the set $\mathit{NS}_2$ as the set of tripartite boxes that correspond to convex combinations of bipartite no-signaling boxes, i.e., that admit a decomposition of the form (\ref{svet}) with $P_\lambda(ab|xy)$ in (\ref{AB/C}) being restricted to be no-signaling (and similarly for other partitions of the parties). This represent the set of tripartite boxes that are in essence only bipartite. In the case of binary inputs and outputs, extremal no-signaling boxes of the form $P_\lambda(ab|xy)$ correspond either to local deterministic boxes or PR boxes. The set $\mathit{NS}_2$ thus corresponds to the boxes that can be simulated using shared randomness and a single PR box shared between any two parties. 

We clearly have the inclusions $L\subseteq \mathit{NS}_2\subseteq\mathit{US}_2\subseteq \mathit{KS}_2\subseteq S_2$ (furthermore, these inclusions are strict, see \cite{inprep}). Each of these sets corresponds to a polytope that can be characterized using linear programming, making it easy to determine whether a given box belongs to any of them.

\subsection{Nonlocality of the extremal boxes}

In Table~\ref{tnoise}, we computed the resistance to white noise of extremal boxes according to all these different notions of nonlocality. That is, we computed (using linear programming) the minimal value $q$ such that the noisy box characterized by the probability distribution $(1-q) P(abc|xyz) +q/8$ belongs to any one of the sets. The series of inclusions $L\subseteq \mathit{NS}_2\subseteq\mathit{US}_2\subseteq \mathit{KS}_2\subseteq S_2$ implies that for a given type of box (corresponding to a given row of Table~\ref{tnoise}) the corresponding noise resistances can only decrease.

Note that the extremal boxes in Tables~\ref{tvertphys} and \ref{tvertcomp} have been ordered according to the noise resistance computed in Table~\ref{tnoise}. That is, boxes have been first ordered in ascending order with respect to their noise resistance for $S_2$. Boxes with the same noise resistance for $S_2$ have been ordered with respect to their noise resistance for $\mathit{KS}_2$, and so on. (The relative ordering between boxes with the same noise resistance according to all notions is arbitrary).

From Table~\ref{tnoise}, one obtains the following.
\begin{itemize}
\item \emph{$L$ boxes (local boxes)}: as noted earlier, boxes 1 are the only local extremal boxes and they correspond to the vertices of the local polytope $L$. All other extremal boxes are nonlocal.

\item \emph{$\mathit{NS}_2$ boxes (bipartite no-signaling boxes)}: as noted above, this class comprises boxes 1 (the deterministic ones) and boxes 2 (the PR-boxes). These two types of boxes define the vertices of the bipartite no-signaling polytope $\mathit{NS}_2$. All other extremal boxes are genuine tripartite no-signaling boxes in the sense that they can only be reproduced using no-signaling resources shared between all three parties; they are the analogous of the quantum correlations that can be obtained only by measuring genuinely three-partite entangled states. Not all genuinely tripartite no-signaling boxes, however, are 3-way nonlocal, as shown below.

\item \emph{$\mathit{US}_2$ boxes (two-way nonlocal boxes)}: this class comprises boxes 1 to 5. All these boxes can be reproduced using classical communication between only two of the parties, even if inputs are given in an arbitrary sequence unknown to the parties. Consider box 3 for instance who is defined by the relations
\be 
\hat a_0+\hat b_y=1,\quad	\hat a_1+\hat c_0=1,\quad \hat a_1+\hat b_y+\hat c_1=y,
\ee
all other correlation terms being uniformly random. 
Here is a model reproducing it involving only communication between Alice and Bob. The model uses two shared random variables $\lambda_0$ and $\lambda_1$ both taking the values $0$ or $1$ with equal probability. 
Charles produces his outputs according to $\hat c_z=\lambda_z$. If Alice receives her input first, she outputs $\hat a_0=\lambda_0+\lambda_1$ or $\hat a_1=\lambda_0$ and Bob outputs $\hat b_y=\lambda_0+\lambda_1$ if Alice's input is $x=0$, and $\hat b_y=\lambda_0+\lambda_1+y $ if Alice's input is $x=1$. If Bob receives his input first, he outputs $\hat b_y=\lambda_0+\lambda_1+y$ and Alice outputs $\hat a_0=\lambda_0+\lambda_1+y$ or $\hat a_1=\lambda_0$. It is easy to see that this model correctly reproduces box 3. All extremal boxes that do not belong to $\mathit{US}_2$ manifest genuine tripartite non-locality, in the sense, that their simulation requires communication between all three parties in at least one experimental situation (corresponding to inputs given in an unkown sequence). 

\item \emph{$\mathit{KS}_2$ boxes}: this class comprises boxes 1 to 8.

\item \emph{$S_2$ boxes}: this class comprises boxes 1 to 12.
\end{itemize}

Finally, note that the boxes that are the most nonlocal according to all notions of 2-way nonlocality are boxes 45 and 46. 

\section{Interconversion between boxes}
\label{sec:inter}

The classification through nonlocality that we have just presented exhibits a rich structure, which we may not have fully exploited in the discussion above. Moreover, it is not the only possible approach. A different classification, for instance, may be based on the possibility of \textit{simulating} some boxes using other ones. We do not attempt a systematic study here, but want to point out that this classification will look astonishingly different from the one based on non-locality.

Indeed, consider just the question of whether a given tripartite box can be simulated by sharing any number of \textit{bipartite PR-boxes} between each pair of parties. Boxes 44, 45, and 46, which are the most non-local according to the previous criterion, are easily simulated using at most one PR-box between each pair of parties \cite{blmppr}. (More generally, it was shown in \cite{bp05}, that any $n$-partite full correlation box can be simulated with PR boxes.) On the contrary, box 4 cannot be simulated even by sharing infinitely many PR-box between the pairs \cite{bp05,fff}; but it is pretty weak on the non-locality scale and, as we shall mention below, it does not violate maximally any of the Bell's inequalities.

\section{Extremal points and Bell-type inequalities}
\label{sec:bell}

\subsection{Overview of the inequalities}
In this section, we focus in greater detail on how extremal boxes differ from local correlations. For three parties and binary inputs and outputs, the local set was fully characterized by Pitowsky and Svozil \cite{pito} and \'Sliwa \cite{sliwa}. The local polytope has $53856$ facets defining $46$ different classes of inequalities that are inequivalent under relabeling of parties, inputs, and outputs. Only a few of these inequalities have been studied more or less thoroughly. \textit{Inequality 1} (we follow \'Sliwa's numbering) is a trivial facet, it corresponds to the condition that probabilities must be comprised between 0 and 1; obviously, no point can violate it. \textit{Inequality 2} is the Mermin inequality (\ref{merminref}), \textit{inequality 4} is the CHSH inequality \cite{chsh}, while \textit{inequality 10} is the GYNI inequality (\ref{gyniineq}). 

\subsection{Violations by the extremal points}

One of the most obvious questions to address is: for any inequality, find the extremal points that return the highest no-signaling violation. The boxes that violate maximally each inequality\footnote[7]{Of course, there is no guarantee that the representative point written in Tables \ref{tvertphys} and \ref{tvertcomp} is the optimal one for the representative inequalities as written in \cite{sliwa}.} are given in Table \ref{tab:sliwac}. Some remarks on this table:
\begin{itemize}
\item Boxes of class 2 (i.e.~PR boxes, the least non-local ones according to the criteria of Section \ref{sec:nonloc}) violate maximally 28 of the inequalities, and for 14 inequalities they are the only ones that do so.
\item Boxes 3, 4, 5, 10, 15, 18, 26, 30, 31, 32 and 33 do not violate any inequality maximally.
\item Boxes 46, which, as we have seen above, are related to the GHZ argument, violate maximally only the inequalities 2 (Mermin), 22 and 33. Boxes 45, which are as non-local as boxes 46 according to Table~\ref{tnoise}, violate a total of 8 inequalities.
\end{itemize}

For a given extremal point $P(abc|xyz)$, we can also compute for each class of Bell inequalities, the minimal amount of white noise $q$ such that the noisy point $(1-q) P(abc|xyz) +q/8$ no longer violates any inequality in the class. Tables \ref{tab:sliwaa} and \ref{tab:sliwab} report these value for each of the 46 extreme points. 
A positive value $q>0$ indicates that the corresponding extremal point violates some inequality in the class (i.e. we must add some noise $q>0$ so that it ceases to violate the inequalities). Among the extremal points that do not violate any inequality in a class, we can distinguish between those that lie on the border of the region defined by these inequalities (i.e. those that reach the local bound of at least one inequality) and those that belong to the interior of this region (i.e. those that do not even reach the local bound of any of the inequalities). We distinguish these two situations by reporting values $q=0$ and $q<0$, respectively. Some information that can be extracted from these tables is as follows.
\begin{itemize}
\item By reading the tables column by column: for a \textit{given inequality}, the boxes that give the largest violation, i.e., that have the largest resistance to noise, correspond to those listed in Table \ref{tab:sliwac}. 
\item By reading the table line by line: for 29 out of the 45 nonlocal boxes, the best resistance to noise is obtained with inequality number 4, i.e., CHSH. It comes as a surprise that an inequality, which is effectively tailored for two parties, is single out so markedly in a three-partite scenario; we recall that a similar situation is encountered for the resistance to noise when one stays in the bipartite case but increases the number of outcomes \cite{acindurt}.
\end{itemize}

\section{Conclusions}

We have studied the no-signaling polytope corresponding to the Bell scenario with three parties, two inputs, and two outputs. Here we summarize some of the properties that we discussed:
\begin{itemize}
\item[-] The polytope has 53856 extremal points, belonging to 46 non-equivalent classes upon relabeling of the parties, inputs, and outputs (Tables \ref{tvertphys} and \ref{tvertcomp}).
\item[-] The extremal points can be classified according to their nonlocality, measured in different ways (Table \ref{tnoise}). In this sense, the most non-local points appear to be boxes 45 and 46, the latter being related to the GHZ correlations in quantum physics. We also mentioned another criterion, based on interconversion of resources, and showed that it would lead to a very different classification.
\item[-] In this scenario, there are also 53856 Bell-type inequalities (i.e. facets of the local polytope) belonging to 46 inequivalent classes, but we have not found any simple one-to-one correspondence with the classes of points. In fact, boxes 2 (the least non-local of all) violate maximally many of the inequalities; while several boxes do not violate maximally any of the inequalities (Table~\ref{tab:sliwac}).
\end{itemize}

Much more information can certainly be extracted from the lists of points and from their properties presented here. For instance, a detailed study of the properties of each no-signaling point is lacking. It would also be interesting to understand how many inequivalent classes of boxes they are with respect to interconversions (for instance, boxes 2, 44, 45, 46 all belong to the same class because from boxes 2 we can obtain boxes 44, 45, 46, and conversely from boxes 44, 45, 46 we can obtain boxes 2). Finally, another open possibility is that some of the classes of extremal points may be ``irrelevant" for quantum correlations, in the sense that all quantum correlations could be decomposed in extremal no-signaling points without ever using any point belonging to those classes.

\vspace*{-0.5em}
\paragraph{Acknowledgments.}
The list of extremal points was generated some years ago and we acknowledge discussions along these years with several colleagues, among whom Jonathan Barrett, Nicolas Gisin, Thinh Phuc Le, Serge Massar, Sandu Popescu, David Roberts. This work was supported by the National Research Foundation and the Ministry of Education, Singapore, the Swiss NCCR Quantum Photonics, the European ERC-AG QORE, and the Brussels-Capital region through a BB2B grant.

\newpage

\begin{longtable}{|c|c@{\;}cc@{\;}cc@{\;}cc@{\;}c@{\;}c@{\;}cc@{\;}c@{\;}c@{\;}cc@{\;}c@{\;}c@{\;}cc@{\;}c@{\;}c@{\;}c@{\;}c@{\;}c@{\;}c@{\;}r|r|}
\caption{Extremal boxes of the tripartite no-signaling polytope. A representative is given for each equivalence class of boxes under relabelling of parties, inputs, and outputs. $n_e$ denote the number of representatives in each class.}\label{tvertphys}\\
\hline
& \multicolumn{2}{c}{$\langle A_x\rangle$}& \multicolumn{2}{c}{$\langle B_y\rangle$} & \multicolumn{2}{c}{$\langle C_z\rangle$} & \multicolumn{4}{c}{$\langle A_xB_y\rangle$} & \multicolumn{4}{c}{$\langle A_xC_z\rangle$} &\multicolumn{4}{c}{$\langle B_yC_z\rangle$} &\multicolumn{8}{c|}{$\langle A_xB_yC_z\rangle$} &
\\
\#&\scriptsize 0&\scriptsize 1&\scriptsize 0&\scriptsize 1&\scriptsize 0&\scriptsize 1&\scriptsize 00&\scriptsize 01&\scriptsize 10& \scriptsize 11&\scriptsize 00&\scriptsize 01&\scriptsize 10&\scriptsize 11&\scriptsize 00&\scriptsize 01&\scriptsize 10&\scriptsize 11&\scriptsize 000&\scriptsize 001&\scriptsize 010&\scriptsize 011&\scriptsize 100&\scriptsize 101& \scriptsize 110&\scriptsize 111&$n_e$\\
\hline
\endfirsthead
\caption{continued.}\\
\hline
& \multicolumn{2}{c}{$\langle A_x\rangle$}& \multicolumn{2}{c}{$\langle B_y\rangle$} & \multicolumn{2}{c}{$\langle C_z\rangle$} & \multicolumn{4}{c}{$\langle A_xB_y\rangle$} & \multicolumn{4}{c}{$\langle A_xC_z\rangle$} &\multicolumn{4}{c}{$\langle B_yC_z\rangle$} &\multicolumn{8}{c|}{$\langle A_xB_yC_z\rangle$} &
\\
\#&\scriptsize 0&\scriptsize 1&\scriptsize 0&\scriptsize 1&\scriptsize 0&\scriptsize 1&\scriptsize 00&\scriptsize 01&\scriptsize 10& \scriptsize 11&\scriptsize 00&\scriptsize 01&\scriptsize 10&\scriptsize 11&\scriptsize 00&\scriptsize 01&\scriptsize 10&\scriptsize 11&\scriptsize 000&\scriptsize 001&\scriptsize 010&\scriptsize 011&\scriptsize 100&\scriptsize 101& \scriptsize 110&\scriptsize 111&$n_e$\\
\hline
\endhead
  \multicolumn{12}{l}{{Continued on the next page\ldots}} \\
\endfoot
  \hline
\endlastfoot

1&1&1&1&1&1&1&1&1&1&1&1&1&1&1&1&1&1&1&1&1&1&1&1&1&1&1&64\\[.5em]
2&1&1&0&0&0&0&0&0&0&0&0&0&0&0&1&1&1&-1&1&1&1&-1&1&1&1&-1&96\\[.5em]
3&0&0&0&0&0&0&1&1&0&0&0&0&1&0&0&0&0&0&0&0&0&0&0&1&0&-1&384\\[.5em]
4&0&0&0&0&0&0&1&0&0&0&0&0&1&0&0&0&0&1&0&0&1&0&0&-1&0&0&128\\[.5em]
5&0&0&0&0&0&0&1&0&0&1&1&0&0&0&1&0&0&0&0&0&0&1&0&-1&0&0&384\\[.5em]
6&0&0&0&0&0&0&1&1&0&0&0&0&0&0&0&0&0&0&0&0&0&0&1&1&1&-1&384\\[.5em]
7&0&0&0&0&0&0&1&0&0&1&0&0&0&0&0&0&0&0&0&0&1&1&1&-1&0&0&192\\[.5em]
8&0&0&0&0&0&0&1&1&0&0&1&0&0&0&1&0&1&0&0&0&0&0&0&1&0&-1&192\\
\hline
\end{longtable}

\clearpage

\newpage

\begin{landscape}

\begin{longtable}[r]{|r|p{4em}@{\;\;}p{4em}@{\;\;}p{4em}@{\;\;}p{8em}@{\;\;\;}p{8em}@{\;\;\;}p{8em}@{\;\;\;}p{11em}|}
\caption{Extremal boxes in the computer scientist notation.}\label{tvertcomp}\\

\hline \# & $\langle \hat{A}_x\rangle$ & $\langle \hat{B}_y\rangle$ & $\langle \hat{C}_z\rangle$ & $\langle \hat{A}_x+\hat{B}_y\rangle$ & $\langle \hat{A}_x+\hat{C}_z\rangle$ & $\langle \hat{B}_y+\hat{C}_z\rangle$ & $\langle \hat{A}_x+\hat{B}_y+\hat{C}_z\rangle$\\ \hline
& & & & & & &\\[-0.25em]
\endfirsthead
\caption{continued.}\\
\hline \# & $\langle \hat{A}_x\rangle$ & $\langle \hat{B}_y\rangle$ & $\langle \hat{C}_z\rangle$ & $\langle \hat{A}_x+\hat{B}_y\rangle$ & $\langle \hat{A}_x+\hat{C}_z\rangle$ & $\langle \hat{B}_y+\hat{C}_z\rangle$ & $\langle \hat{A}_x+\hat{B}_y+\hat{C}_z\rangle$\\ \hline& & & & & & &\\[-0.25em]
\endhead

  \multicolumn{8}{l}{{Continued on the next page\ldots}} \\
\endfoot

  \hline
\endlastfoot


1&$ 0  $&$ 0  $&$ 0  $&$ 0  $&$ 0  $&$ 0  $&$ 0  $ \\[.5em]
2&$ 0  $&$ \frac{1}{2}  $&$ \frac{1}{2}  $&$ \frac{1}{2}  $&$ \frac{1}{2}  $&$ yz  $&$ yz $ \\[.5em]
3&$ \frac{1}{2}  $&$ \frac{1}{2}  $&$ \frac{1}{2}  $&$ \frac{1}{2}x  $&$ \frac{1}{2}(1-x+xz)$&$ \frac{1}{2}  $&$ \frac{1}{2}(1-xz)+xyz $ \\[.5em]
4&$ \frac{1}{2}  $&$ \frac{1}{2}  $&$ \frac{1}{2}  $&$ \frac{1}{2}(x+y-xy)$&$ \frac{1}{2}(1-x+xz) $&$ \frac{1}{2}(1-yz)  $&$ \frac{1}{2}(1-y+xy+xz+yz)-xyz $ \\[.5em]
5&$ \frac{1}{2}  $&$ \frac{1}{2}  $&$ \frac{1}{2}  $&$ \frac{1}{2}x+\frac{1}{2}y-xy  $&$ \frac{1}{2}x+\frac{1}{2}z-\frac{1}{2}xz  $&$ \frac{1}{2}y+\frac{1}{2}z-\frac{1}{2}yz  $&$ \frac{1}{2}+\frac{1}{2}xz-\frac{1}{2}yz $ \\[.5em]
6&$ \frac{1}{2}  $&$ \frac{1}{2}  $&$ \frac{1}{2}  $&$ \frac{1}{2}x  $&$ \frac{1}{2}  $&$ \frac{1}{2}  $&$ \frac{1}{2}-\frac{1}{2}x+xyz $ \\[.5em]
7&$ \frac{1}{2}  $&$ \frac{1}{2}  $&$ \frac{1}{2}  $&$ \frac{1}{2}x+\frac{1}{2}y-xy  $&$ \frac{1}{2}  $&$ \frac{1}{2}  $&$ \frac{1}{2}-\frac{1}{2}x-\frac{1}{2}y+xy+xz-xyz $ \\[.5em]
8&$ \frac{1}{2}  $&$ \frac{1}{2}  $&$ \frac{1}{2}  $&$ \frac{1}{2}x  $&$ \frac{1}{2}x+\frac{1}{2}z-\frac{1}{2}xz  $&$ \frac{1}{2}z  $&$ \frac{1}{2}-\frac{1}{2}xz+xyz $ \\
\end{longtable}
\clearpage
\end{landscape}

\newpage
\begin{longtable}{|c|ccccc|}
\caption{Amount of noise that can be tolerated before entering in the local ($L$), bipartite no-signalling ($\mathit{NS}_2$), Svetlichny with unknown input sequence ($\mathit{US}_2$), Svetlichny with fixed input sequence ($\mathit{KS}_2$), and original Svetlichny ($S_2$) polytopes. The asterisk denote the highest resistance in each column.}\label{tnoise}\\
  \hline
  \# & $L$ & $\mathit{NS}_2$ & $\mathit{US}_2$ & $\mathit{KS}_2$ & $\mathit{S}_2$ \\
  \hline
\endfirsthead
\caption{continued.}\\
  \hline
  \# & $L$ & $\mathit{NS}_2$ & $\mathit{US}_2$ & $\mathit{KS}_2$ & $\mathit{S}_2$ \\
  \hline
\endhead
  \multicolumn{6}{l}{{Continued on the next page\ldots}} \\
\endfoot
  \hline
\endlastfoot
1 & $0$& $0$& $0$& $0$& $0$ \\ 
2 & $2/3^*$& $0$& $0$& $0$& $0$ \\ 
3 & $1/2$& $1/3$& $0$& $0$& $0$ \\ 
4 & $2/5$& $2/5$& $0$& $0$& $0$ \\ 
5 & $1/2$& $2/5$& $0$& $0$& $0$ \\ 
6 & $1/2$& $1/3$& $1/3$& $0$& $0$ \\ 
7 & $1/2$& $1/3$& $1/3$& $0$& $0$ \\ 
8 & $1/2$& $2/5$& $2/5$& $0$& $0$ \\
\hline
\end{longtable}
\clearpage

\begin{longtable}{|c|c||c|c|}\caption{Boxes attaining the maximal violation of each Bell-type inequality. The inequalities are listed from 1 to 46, following \cite{sliwa}. Boxes are numbered according to Tables \ref{tvertphys} and \ref{tvertcomp}.}\label{tab:sliwac}\\
  \hline
Ineq. & Extremal boxes & Ineq. & Extremal boxes \\
  \hline
1 & none & 24 & 29\\[.2em]
2 & 2, 21, 22, 34, 44, 46 & 25 & 25\\[.2em]
3 & 2, 7, 9, 16, 19, 20, 23, 35, 37, 42, 44, 45 & 26 & 29\\[.2em]
4 & 2 & 27 & 25, 28, 29\\[.2em]
5 & 2 & 28 & 2\\[.2em]
6 & 2 & 29 & 2\\[.2em]
7 & 44 & 30 & 2\\[.2em]
8 & 2, 7, 11, 13, 16, 19, 21, 42, 44 & 31 & 19, 37, 42, 44, 45\\[.2em]
9 & $\begin{array}{l}2, 6, 12, 14, 17, 19, 23, 24, 27,\\ 28, 35, 37, 38, 41, 42, 44, 45\end{array}$ & 32 & 42\\[.8em]
10 & 25, 29 & 33 & 21, 44, 46\\[.2em]
11 & 2, 41, 45 & 34 & 41\\[.2em]
12 & 2 & 35 & 41, 45\\[.2em]
13 & 2, 42 & 36 & 2\\[.2em]
14 & 2 & 37 & 2\\[.2em]
15 & 2, 40 & 38 & 2\\[.2em]
16 & 2, 19, 21, 27, 28, 42, 44 & 39 & 46\\[.2em]
17 & 2, 8, 12, 19, 23 & 40 & 40, 42, 45\\[.2em]
18 & 2, 41 & 41 & 2, 19, 44\\[.2em]
19 & 2, 40, 42 & 42 & 2, 21, 29, 44\\[.2em]
20 & 2 & 43 & 2, 19, 25, 28, 35, 37, 38, 42, 43, 44, 45\\[.2em]
21 & 43 & 44 & 2\\[.2em]
22 & 2, 19, 21, 27, 34, 36, 39, 41, 44, 46 & 45 & 2\\[.2em]
23 & 25, 45 & 46 & 25\\[.2em]
\hline
\end{longtable}

\newpage
\begin{landscape}
\begin{longtable}{|c|ccccccccccccccccccccccc|}
\caption{For each class of boxes, maximum amount of noise tolerated before ceasing to violate a given family of Bell-type inequalities. Each line corresponds to a box and each column to an inequality. The inequality numbers refers to the ones chosen in \cite{sliwa}. Table for inequalities 1 to 23.}\label{tab:sliwaa}\\
  \hline
  \# & 1 & 2 & 3 & 4 & 5 & 6 & 7 & 8 & 9 & 10 & 11 & 12 & 13 & 14 & 15 & 16 & 17 & 18 & 19 & 20 & 21 & 22 & 23\\[.5em]
  \hline
\endfirsthead
\caption{continued.}\\
  \hline
  \# & 1 & 2 & 3 & 4 & 5 & 6 & 7 & 8 & 9 & 10 & 11 & 12 & 13 & 14 & 15 & 16 & 17 & 18 & 19 & 20 & 21 & 22 & 23\\[.5em]
  \hline
\endhead
  \multicolumn{24}{l}{{Continued on the next page\ldots}} \\
\endfoot
  \hline
\endlastfoot
1 & 0 & 0 & 0 & 0 & 0 & 0 & 0 & 0 & 0 & 0 & 0 & 0 & 0 & 0 & 0 & 0 & 0 & 0 & 0 & 0 & 0 & 0 & 0\\[.5em]
2 & 0 & $\frac{1}{2}$ & $\frac{1}{2}$ & $\frac{2}{3}$ & $\frac{4}{7}$ & $\frac{4}{7}$ & $\frac{1}{2}$ & $\frac{1}{2}$ & $\frac{1}{2}$ & 0 & $\frac{1}{2}$ & $\frac{1}{2}$ & $\frac{1}{2}$ & $\frac{1}{2}$ & $\frac{1}{2}$ & $\frac{1}{2}$ & $\frac{1}{2}$ & $\frac{1}{2}$ & $\frac{1}{2}$ & $\frac{3}{5}$ & $\frac{1}{2}$ & $\frac{1}{2}$ & $\frac{1}{3}$\\[.5em]
3 & 0 & -1 & 0 & $\frac{1}{2}$ & $\frac{2}{5}$ & $\frac{2}{5}$ & 0 & 0 & $\frac{1}{3}$ & 0 & 0 & $\frac{1}{3}$ & $\frac{1}{3}$ & $\frac{1}{3}$ & $\frac{1}{3}$ & 0 & $\frac{1}{3}$ & $\frac{1}{3}$ & $\frac{1}{3}$ & $\frac{1}{5}$ & 0 & 0 & $\frac{1}{5}$\\[.5em]
4 & 0 & -1 & 0 & 0 & $\frac{2}{5}$ & $\frac{2}{5}$ & 0 & 0 & 0 & 0 & 0 & $\frac{1}{3}$ & 0 & 0 & 0 & 0 & 0 & $\frac{1}{3}$ & 0 & $\frac{1}{5}$ & 0 & 0 & $\frac{1}{5}$\\[.5em]
5 & 0 & 0 & 0 & $\frac{1}{2}$ & $\frac{2}{5}$ & $\frac{2}{5}$ & 0 & $\frac{1}{3}$ & 0 & $\frac{1}{3}$ & $\frac{1}{3}$ & $\frac{1}{3}$ & 0 & $\frac{1}{3}$ & $\frac{1}{3}$ & 0 & 0 & $\frac{1}{3}$ & $\frac{1}{3}$ & $\frac{1}{5}$ & $\frac{1}{3}$ & $\frac{1}{3}$ & $\frac{1}{5}$\\[.5em]
6 & 0 & 0 & 0 & $\frac{1}{2}$ & $\frac{2}{5}$ & $\frac{2}{5}$ & $\frac{1}{3}$ & $\frac{1}{3}$ & $\frac{1}{2}$ & 0 & $\frac{1}{3}$ & 0 & $\frac{1}{3}$ & $\frac{1}{3}$ & $\frac{1}{3}$ & $\frac{1}{3}$ & $\frac{1}{3}$ & $\frac{1}{3}$ & $\frac{1}{3}$ & $\frac{1}{3}$ & $\frac{1}{3}$ & $\frac{1}{3}$ & $\frac{1}{3}$\\[.5em]
7 & 0 & 0 & $\frac{1}{2}$ & $\frac{1}{2}$ & $\frac{2}{5}$ & $\frac{2}{5}$ & $\frac{1}{3}$ & $\frac{1}{2}$ & $\frac{1}{3}$ & 0 & $\frac{1}{3}$ & $\frac{1}{3}$ & $\frac{1}{3}$ & 0 & 0 & $\frac{1}{3}$ & $\frac{1}{3}$ & $\frac{1}{3}$ & $\frac{1}{3}$ & $\frac{1}{3}$ & $\frac{1}{3}$ & 0 & $\frac{1}{5}$\\
\hline

\end{longtable}
\newpage

\begin{longtable}{|c|ccccccccccccccccccccccc|}
\caption{As Table 5, but for inequalities 24 to 46.}\label{tab:sliwab}\\
  \hline
  \# & 24 & 25 & 26 & 27 & 28 & 29 & 30 & 31 & 32 & 33 & 34 & 35 & 36 & 37 & 38 & 39 & 40 & 41 & 42 & 43 & 44 & 45 & 46\\[.5em]
  \hline
\endfirsthead
\caption{continued.}\\
  \hline
  \# & 24 & 25 & 26 & 27 & 28 & 29 & 30 & 31 & 32 & 33 & 34 & 35 & 36 & 37 & 38 & 39 & 40 & 41 & 42 & 43 & 44 & 45 & 46\\[.5em]
  \hline
\endhead
  \multicolumn{24}{l}{{Continued on the next page\ldots}} \\
\endfoot
  \hline
\endlastfoot
1 & 0 & 0 & 0 & 0 & 0 & 0 & 0 & 0 & 0 & 0 & 0 & 0 & 0 & 0 & 0 & 0 & 0 & 0 & 0 & 0 & 0 & 0 & 0\\[.5em]
2 & $\frac{4}{9}$ & $\frac{4}{9}$ & $\frac{4}{9}$ & $\frac{4}{9}$ & $\frac{4}{7}$ & $\frac{4}{7}$ & $\frac{4}{7}$ & $\frac{2}{5}$ & $\frac{2}{5}$ & $\frac{2}{5}$ & $\frac{2}{5}$ & $\frac{2}{5}$ & $\frac{4}{7}$ & $\frac{4}{7}$ & $\frac{4}{7}$ & $\frac{2}{5}$ & $\frac{2}{5}$ & $\frac{8}{15}$ & $\frac{1}{2}$ & $\frac{1}{2}$ & $\frac{3}{5}$ & $\frac{3}{5}$ & $\frac{4}{9}$\\[.5em]
3 & 0 & $\frac{2}{7}$ & 0 & $\frac{2}{7}$ & 0 & $\frac{1}{4}$ & $\frac{1}{4}$ & $\frac{1}{4}$ & $\frac{1}{4}$ & 0 & $\frac{1}{4}$ & $\frac{1}{4}$ & 0 & $\frac{1}{4}$ & $\frac{1}{4}$ & 0 & $\frac{1}{4}$ & $\frac{2}{9}$ & 0 & $\frac{1}{5}$ & $\frac{1}{5}$ & $\frac{1}{5}$ & $\frac{1}{6}$\\[.5em]
4 & 0 & $\frac{2}{7}$ & 0 & $\frac{2}{7}$ & 0 & $\frac{1}{4}$ & $\frac{1}{4}$ & $\frac{1}{4}$ & $\frac{1}{4}$ & 0 & $\frac{1}{4}$ & $\frac{1}{4}$ & 0 & $\frac{1}{4}$ & $\frac{1}{4}$ & 0 & $\frac{1}{4}$ & 0 & 0 & $\frac{1}{5}$ & $\frac{1}{5}$ & $\frac{1}{5}$ & $\frac{2}{7}$\\[.5em]
5 & $\frac{2}{7}$ & $\frac{2}{7}$ & $\frac{4}{9}$ & $\frac{2}{7}$ & $\frac{1}{4}$ & $\frac{1}{4}$ & $\frac{1}{4}$ & $\frac{1}{4}$ & $\frac{1}{4}$ & $\frac{1}{4}$ & $\frac{1}{4}$ & $\frac{1}{4}$ & $\frac{1}{4}$ & $\frac{1}{4}$ & $\frac{1}{4}$ & $\frac{1}{4}$ & $\frac{1}{4}$ & 0 & $\frac{1}{3}$ & $\frac{1}{5}$ & $\frac{1}{5}$ & $\frac{1}{5}$ & $\frac{2}{7}$\\[.5em]
6 & $\frac{2}{7}$ & $\frac{2}{7}$ & 0 & $\frac{2}{7}$ & $\frac{1}{4}$ & $\frac{1}{4}$ & $\frac{2}{5}$ & $\frac{1}{4}$ & $\frac{2}{5}$ & $\frac{1}{4}$ & $\frac{2}{5}$ & $\frac{2}{5}$ & $\frac{1}{4}$ & $\frac{2}{5}$ & $\frac{1}{4}$ & $\frac{1}{4}$ & $\frac{1}{4}$ & $\frac{4}{11}$ & $\frac{1}{5}$ & $\frac{1}{3}$ & $\frac{1}{5}$ & $\frac{1}{3}$ & $\frac{2}{7}$\\
\hline

\end{longtable}
\end{landscape}

\end{document}